\documentclass[%
 aip,
 jmp,%
 amsmath,amssymb,
 reprint,%
]{revtex4-2}
\newcommand{\blue}[1]{\textcolor{blue}{#1}}

\usepackage{graphicx,color,mhchem}
\usepackage{dcolumn}
\usepackage{bm}
\usepackage{placeins}
\begin{document}

\title[Tantalum oxidation]{Oxidation of Tantalum Nano-Film by Microwave Exposure}

\author{
Massimiliano Zamengo$^{1,*}$, Lina Grineviciute$^{2}$, Hsin-Hui Huang$^{3,4}$,
Haoran Mu$^{3,4}$, Julianija Nikitina$^{2}$, 
Saulius Juodkazis$^{3,5,6}$, Junko Morikawa$^{1,5,7,*}$
}
\thanks{Correspondence: M.Z. zamengo.m.82d4@m.isct.ac.jp; J.M. morikawa.j.4f50@m.isct.ac.jp}
\affiliation{School of Materials and Chemical Technology, 
The Institute of Science Tokyo, 2-12-1 Ookayama, Meguro-ku, Tokyo 152-8550, Japan}
\affiliation{Center for Physical Sciences and Technology (FTMC), Savanoriu Ave. 231, LT-02300 Vilnius, Lithuania}
\affiliation{Optical Sciences Centre, School of Science, Swinburne University of Technology, Hawthorn, Victoria 3122, Australia}
\affiliation{Melbourne Centre for Nanofabrication (MCN), 151 Wellington Road, Clayton, Victoria 3168, Australia}
\affiliation{World Research Hub (WRH), School of Materials and Chemical Technology, Institute of Science Tokyo, 2-12-1 Ookayama, Meguro-ku, Tokyo 152-8550, Japan}
\affiliation{Laser Research Center, Physics Faculty, Vilnius University, Saul\.{e}tekio Ave. 10, 10223 Vilnius, Lithuania}
\affiliation{Research Center for Autonomous Systems Materialogy (ASMat), Institute of Innovative Research, Institute of Science Tokyo, Yokohama 226-8501, Japan}

\date{\today}

\begin{abstract}
Oxidation and ablation of 200~nm tantalum films were carried out by three routes: (i) femtosecond (fs-)laser direct write, (ii) high-temperature annealing (HTA) in a tube furnace, and (iii) annealing in a 2.45~GHz microwave cavity. Complete conversion of the 200~nm Ta layer into 409~nm of \ce{Ta2O5} required one hour at $600^\circ$C in the furnace, but only minutes at $\sim 50$~W of microwave power. Fs-laser (515~nm/200~fs) oxidation of the Ta nano-film set in at an average single-pulse 
fluence of $\sim 0.1$~J/cm$^2$ under strong pulse-to-pulse overlap (900 pulses per focal spot), 
i.e. within a narrow window bounded from above by the onset of ablation. 
Under microwave annealing, both the cavity resonance frequency and the quality factor $Q$ changed markedly 
at the metal-to-oxide transition, reflecting the collapse of the real and imaginary parts of the permittivity at 2.45~GHz. 
This 
dielectric contrast turns the cavity into a sensor: the oxidation can be followed in real time from the shift of the cavity resonance, providing a non-invasive, in-situ diagnostic tool.
\end{abstract}

\maketitle


\section{Introduction}

Tantalum is a biocompatible metal, well suited as a coating for implantable capsules that house the electro-optical components used in nerve stimulation. Optogenetics is a representative application: visible to near-IR light from an encapsulated light-emitting diode (LED) must be coupled efficiently into an optical probe or waveguide. This imposes a stringent requirement on the package, since the hermetic housing has to incorporate optically transparent windows. Converting selected regions of the Ta coating into transparent \ce{Ta2O5}, with a high refractive index $n\simeq 2.1$ and a bandgap of $\sim 4$~eV, by laser oxidation is one promising route to such windows~\cite{26olt115018}.

The central difficulty is to drive the oxidation by direct laser writing without triggering ablation. Coupling laser energy into a highly reflective metal is intrinsically inefficient, even when the goal is only the moderate surface heating needed to activate oxidation through an Arrhenius-type thermally activated mechanism. The processing window between sufficient heating and the onset of ablation is correspondingly narrow and demands tight control of the laser parameters.

Microwave energy deposition offers a distinct and complementary alternative to laser-based localized heating. In contrast to laser irradiation, microwaves couple volumetrically to the material and can therefore drive oxidation with no risk of surface ablation. The versatility of this coupling is well documented: curing of epoxy resin proceeds very efficiently when dispersed \ce{SiC} nanoparticles act as absorbers~\cite{sicA}, and volumetric microwave heating is exploited in chemical vapor deposition, where it enhances both heating and plasma formation~\cite{mw}. 

Microwave high-temperature annealing (HTA) is already established across a strikingly broad range of materials problems. In semiconductor processing it restores implantation damage: 4H-SiC implanted with Al$^{+}$ and P$^{+}$ has been annealed at $2120^\circ$C~\cite{sic}, and hydrogen activated by a short $\sim 1$~min microwave anneal passivates the contacts of Si solar cells~\cite{polo}. In synthesis, microwaves drive the growth and crystallization of perovskites from solution for photovoltaics~\cite{pero}, accelerate the solvothermal preparation of fluoride Na$_x$MF$_y$ ($M =$ Co, Mn, Fe) electrocatalysts for fuel cells and sodium batteries~\cite{solvo}, and extend to high-pressure synthesis~\cite{jung}. In extractive metallurgy, microwave HTA strips fluorides from the spent carbon cathodes of aluminium electrolysis~\cite{defl} and assists the water leaching and recovery of sodium fluoride and rare-earth elements~\cite{rare}. Because the heat is generated inside the workpiece rather than transported into it, even flexible electronics can be heat-treated without damage~\cite{park}. It is this route of HTA that is explored in the present study.

Microwave-driven solid-state oxidation, in particular, has emerged as a fast, low-thermal-budget alternative to conventional furnace and rapid-thermal oxidation. It offers inverted thermal gradients (core-to-surface rather than surface-to-core), phase-selective coupling, and processing times reduced from hours to minutes. Recent work spans transition-metal-oxide synthesis, ferroelectric and dielectric thin-film processing, polymer-derived ceramic curing, and integrated diagnostics. Gram-scale direct microwave oxidation of \ce{MoS2} into layered $\alpha$-\ce{MoO3} has been demonstrated, with volumetric heating preserving the belt-like morphology and yielding mm-long crystals suitable as the active layer of low-voltage 
memristors~\cite{Wang2026}. Microwave annealing has been used to crystallize sol-gel-derived Pb(Zr$_{0.4}$Ti$_{0.6}$)O$_3$ epitaxial films, improving their polarization properties at reduced effective furnace temperatures~\cite{Ding2025}. Microwave processing of atomic layer deposition (ALD)-grown \ce{HfO2} thin films enhances the sensitivity, hysteresis, and long-term stability of pH sensors~\cite{Cui2023}, while microwave-assisted oxidation curing of polycarbosilane powders enables rapid cross-linking ($<$1~min to $\sim$800$^{\circ}$C) en route to SiC ceramics~\cite{Hwang2024}. To address the long-standing absence of mechanistic in-situ probes, a time-resolved neutron-diffraction method has been developed for tracking microwave-induced solid-state reactions on sub-second timescales~\cite{McFadzean2025}. Taken together, these studies show microwave-driven solid-state oxidation moving from empirical demonstration toward mechanistic and process-engineering maturity, with the defining advantages of the field --- volumetric heating, phase-selective absorption, and a dramatically reduced thermal budget --- now systematically exploitable across functional oxide platforms.

A resonant cavity of volume $V\sim\lambda^3$ concentrates the microwave field and can heat samples of very different kinds at rates exceeding $100^\circ$C/s, through magnetic, dielectric, or Joule mechanisms of energy deposition~\cite{toyota}. A further advantage of the cavity geometry is that the polarization of the field at the sample is fixed by the cavity mode. Catalytic La-Ce-Ni oxides, for instance, can be sintered by microwave heating that proceeds predominantly via the dielectric channel~\cite{Hamashima}. The energy loss per unit volume, which sets the local energy deposition, is~\cite{Hamashima}
\begin{equation}\label{e-h}
    P_d = \frac{1}{2}\left[\varepsilon_0\varepsilon''\omega|E|^2 + \sigma|E|^2 + \mu_0\mu''\omega|H|^2\right],
\end{equation}
\noindent where $\varepsilon_0$ and $\mu_0$ are the vacuum permittivity and permeability, $\varepsilon''$ and $\mu''$ are the imaginary parts of the permittivity and permeability of the material, $E$ and $H$ are the electric and magnetic field strengths inside the material, and $\omega\equiv 2\pi\nu$ is the cyclic frequency of the microwave field. The first term accounts for dielectric losses, the second for Joule heating, and the third for magnetic coupling to the electromagnetic field. For a purely dielectric loss channel, the first term of Eq.~\ref{e-h} can be written as $P_d = \frac{1}{2}\varepsilon_0\varepsilon'\omega\tan\delta\,|E_0|^2$, where $E_0$ is the amplitude of the applied electric field and the loss tangent $\tan\delta = \varepsilon''/\varepsilon'$ is set by the ratio of the imaginary and real parts of the complex permittivity $\varepsilon^* \equiv \varepsilon' + i\varepsilon''$. Placing the sample in a resonant cavity makes the energy deposition considerably more efficient, because the cavity enhances the $E$-field~\cite{toyota}. The field intensity $|E|^2$ is set by the quality factor $Q$, the cavity volume $V$, and the power $P_0$ coupled into the cavity as~\cite{toyota}
\begin{equation}
 |E|^2  = \frac{4QP_0}{\varepsilon_0\omega V},~~~~|H|^2  = \frac{\varepsilon_0}{\mu_0}|E|^2.
\end{equation}
\noindent The sample for HTA 
is usually inserted into a quartz tube on the cavity axis and oriented along either the $E$- or the $H$-field. Surfaces and interfaces in granular materials carry a large $\varepsilon''$ and hence a large $\tan\delta$, so they act as energy deposition sites: heating starts from inside the sample rather than diffusing inwards from a heated exterior, as it would in conventional HTA~\cite{toyota}. Liquid phases generally have a larger $\tan\delta$ than solids, which localizes the energy deposition further during microwave HTA. Pyrolytic carbon is among the strongest microwave absorbers and therefore an excellent susceptor; it does not contaminate the sample, since at high temperature it burns off as \ce{CO2}. The three-dimensional, volumetric nature of the microwave-matter interaction also opens the way to non-contact characterization in modalities ranging from scanning probe to microscopy~\cite{Li}.

Here we explore spatially controlled oxidation of Ta by two routes: direct write with ultrashort laser pulses (515~nm/200~fs), and HTA in a 2.45~GHz microwave cavity. Direct laser writing (DLW) of sub-wavelength oxidized Ta lines proved to operate uncomfortably close to the ablation threshold and is therefore of limited practical interest. Microwave annealing, by contrast, produced oxide of promising quality at only tens of watts and within tens of seconds --- and, as shown below, carries its own in-situ diagnostic.

\section{Experimental: samples and methods}
Tantalum films 200~nm thick were ion-beam sputtered (Cutting Edge Coatings GmbH, Germany) onto fused silica (\ce{SiO2}) substrates, 25~mm-diameter disks of 1~mm thickness. 
HTA was carried out in two ways: (1) in a high-temperature tube furnace (Nabertherm GmbH, Germany) at $550^\circ$C and $600^\circ$C, and (2) in a microwave cavity. The metallic, non-transparent (reflective, $R\sim 1$) Ta film becomes fully transparent after oxidation in the furnace, with the transparency extending into the IR. Upon complete oxidation the 200~nm Ta film converts into $\sim 409$~nm of \ce{Ta2O5}, as determined by fitting the measured transmittance spectra with OptiLayer software. The thickness increase was confirmed with an optical profilometer (Bruker) and is consistent with the volume expansion that accompanies the oxidation of Ta into \ce{Ta2O5}.

A femtosecond laser (Carbide, 40~W, Light Conversion, Lithuania) integrated into a laser machining setup (WOP, Lithuania) was used for nano- and micro-ablation of the Ta film. The second harmonic, $\lambda = 515$~nm, was used to expose the samples with pedestal-free 200~fs pulses at a range of repetition rates, pulse densities, and scanning speeds; the specific parameters are given in the respective figure captions. 
The enthalpy of fusion of Ta is $\sim 31.7$~kJ/mol (melting point $3017^\circ$C) and its enthalpy of vaporization, i.e. the cohesive binding energy, is 735~kJ/mol, or 7.62~eV/atom. The thermal diffusivity of Ta is modest, $\sim 25$~mm$^2$/s at ambient conditions, comparable with that of air ($\sim 22$~mm$^2$/s).

\subsection{Microwave heating at 2.45~GHz}
Microwave heating and dielectric property measurements were carried out in a cylindrical TM$_{010}$-mode resonator tunable from 2.4 to 2.5~GHz (Minamo, Japan). The resonator was connected through a waveguide, a slug tuner, and coaxial cables to a microwave signal generator (MR-2G-100, Ryowa Denshi, Japan); impedance matching and frequency tracking were handled by dedicated software so that resonance was maintained throughout. The resonant frequency $f$~[MHz] and the quality factor $Q$ were extracted by fitting the VNA data with a pseudo-Voigt (Lorentzian plus Gaussian) profile. Temperature was monitored with a radiation thermometer (TMHX-CNE0500-0070E003, Japan Sensor, Japan).
This setup~\cite{Massi} was capable of launching up to 100~W into the central cavity at 2.45~GHz (12.24~cm wavelength, 0.01~meV photon energy, 0.0817~cm$^{-1}$ wavenumber).
The pyrometer was aimed at the outer wall of the \ce{SiO2} tube directly above the sample and was calibrated for the emissivity of \ce{SiO2}. 
The radiation temperature was determined over the 2-to-6.8~$\mu$m band with an InSb sensor, which reads reliably up to $500^\circ$C. The sensing spot on the sample was 3~mm in diameter at 7~cm depth into the cavity measured from the rim (5~mm at 5~cm and 8~mm at 10~cm depth). Monitoring the radiative temperature directly on the sample surface during heating and oxidation was attempted, but proved unreliable because the emissivity changes as metallic Ta converts into \ce{Ta2O5}. The \ce{SiO2} tube was open to room air throughout all experiments, without forced flow.

A sample can be heated when its own broad absorption resonance partially overlaps the cavity resonance. Samples that reflect the incoming energy back along the waveguide towards the magnetron cannot be heated and have to be reduced in size to limit the reflectance. Non-absorbing samples such as glass can still be processed by conventional diffusive heating, using ceramic microwave susceptors such as \ce{SiC} nano- or micro-powders; 
in our case \ce{SiC} powder was used to heat the top surface of a \ce{SiO2} plate.

\begin{figure*}[b]
\centering\includegraphics[width=1\textwidth]{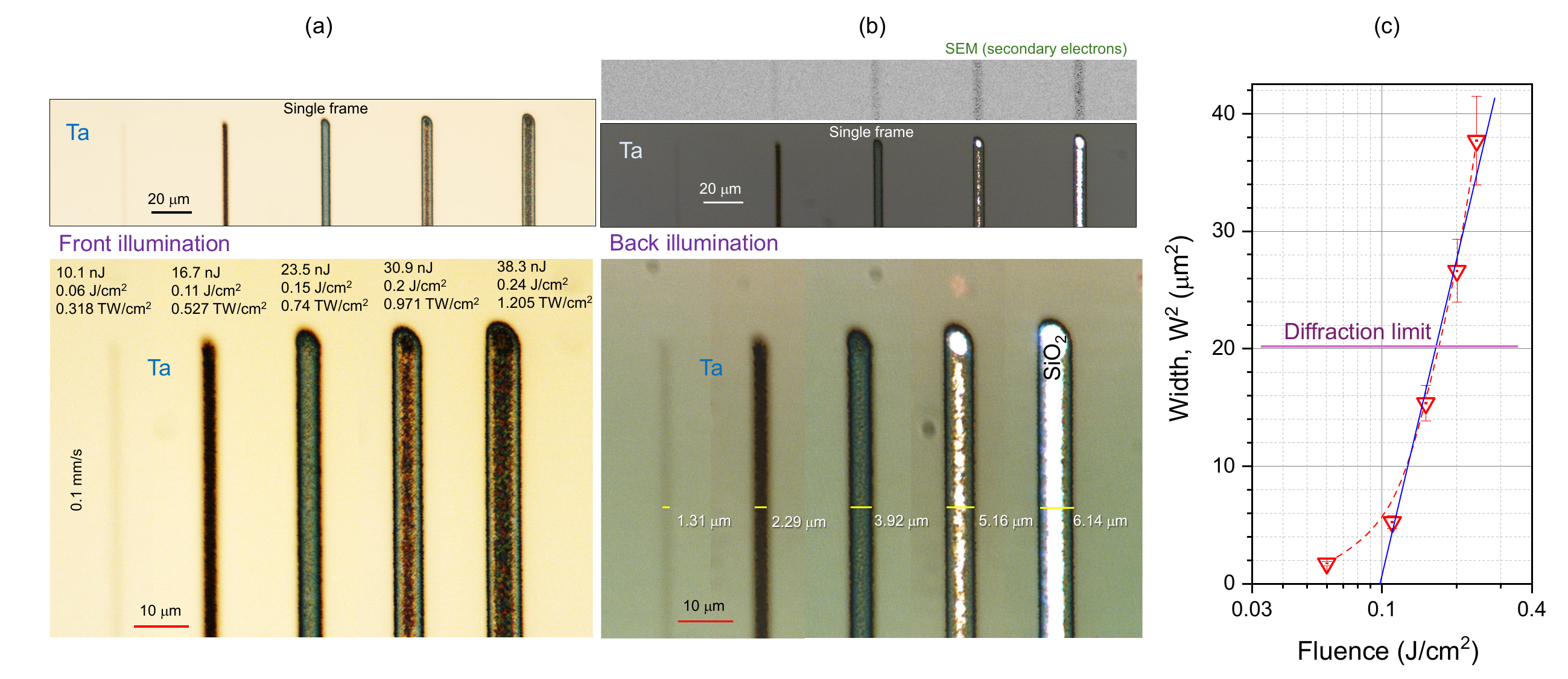}
\caption{\label{f-phot} Optical images of a 200~nm Ta film on a \ce{SiO2} substrate modified/ablated by fs-laser (515~nm/200~fs) exposure, viewed in reflection (a) and transmission (b); the top insets show single-frame images of the same lines. (c) Liu plot of the width $W$ of the modified line (marked) versus the average pulse fluence $F_p$, i.e. $W^2(\lg[F_p])$; error bars are 10\%. The diffraction-limited diameter of the fs-laser writing spot was $\sim 4.5~\mu$m, while the resolution limit of the microscope imaging at $\lambda_M = 500$~nm was $0.61\lambda/NA = 339$~nm for an $NA = 0.9$ objective lens. The ablation threshold in terms of average fluence per pulse was $F_{th} = 0.1$~J/cm$^2$ (0.2~J/cm$^2$ in peak fluence) for $N = 900$ pulses.}
\end{figure*}

A numerical model (COMSOL Multiphysics 6.3) was set up for a quantitative comparison between the experimental conditions, the sample geometry, and the electromagnetic (EM) energy distribution inside the cavity. 

\section{Results}
\subsection{Femtosecond laser surface modification}
Surface ablation and oxidation of the 200~nm Ta film on \ce{SiO2} were carried out with 515~nm/200~fs pulses in constant-density mode, in which the number of pulses per millimetre of scan, $D$~[mm$^{-1}$], is held fixed: when the scan speed $v_s$ is changed, the pulse picking is adjusted accordingly. The effective repetition rate is then $f_{ef} = Dv_s$, which sets the dwell time between adjacent pulses, $t_{dw} = 1/f_{ef}$. Scan speeds of $v_s = (10-100)~\mu$m/s corresponded to $t_{dw} = (500-50)~\mu$s, with a large overlap of $N = 900$ pulses per focal spot of diameter $2r = 1.22\lambda/NA = 4.5~\mu$m. The average fluence per pulse was kept low, $F_p = (0.1-0.2)$~J/cm$^2$, well below the $\sim 0.33$~J/cm$^2$ single-pulse ablation threshold estimated for the same focusing conditions (Fig.~\ref{f-phot}). This regime allowed heat accumulation, diffusion, and oxidation to be explored through the multiplication of structural defects and the accompanying local change of the permittivity $\varepsilon^*\equiv (n^*)^2$, equivalently of the complex refractive index $n^*$.

Controlled modification without apparent ablation was observed only where the width of the modified line fell below the diffraction limit, $\sim 0.61\lambda/NA$ (Fig.~\ref{f-phot}(c)). In the Liu plot, $\mathit{Diameter}^2\propto\lg(\mathit{Fluence})$, this modification lay on the same slope as the ablation branch --- i.e. no separate threshold could be resolved --- and extrapolated to $0.1$~J/cm$^2$. Controlled oxidation by fs-laser direct writing with strongly overlapping exposures at effective repetition rates of 2-20~kHz was therefore not achievable, since ablation and ripple formation set in before oxidation (see details in Ref.~\cite{26olt115018}). Only the high-repetition burst mode delivered controlled oxidation of Ta in direct-write mode~\cite{26arX}; the strong heat accumulation that the burst mode provides proved essential for oxidizing the film without ablating it.

\subsection{Microwave heating/oxidation}
\begin{figure*}[b!]
\centering\includegraphics[width=0.99\textwidth]{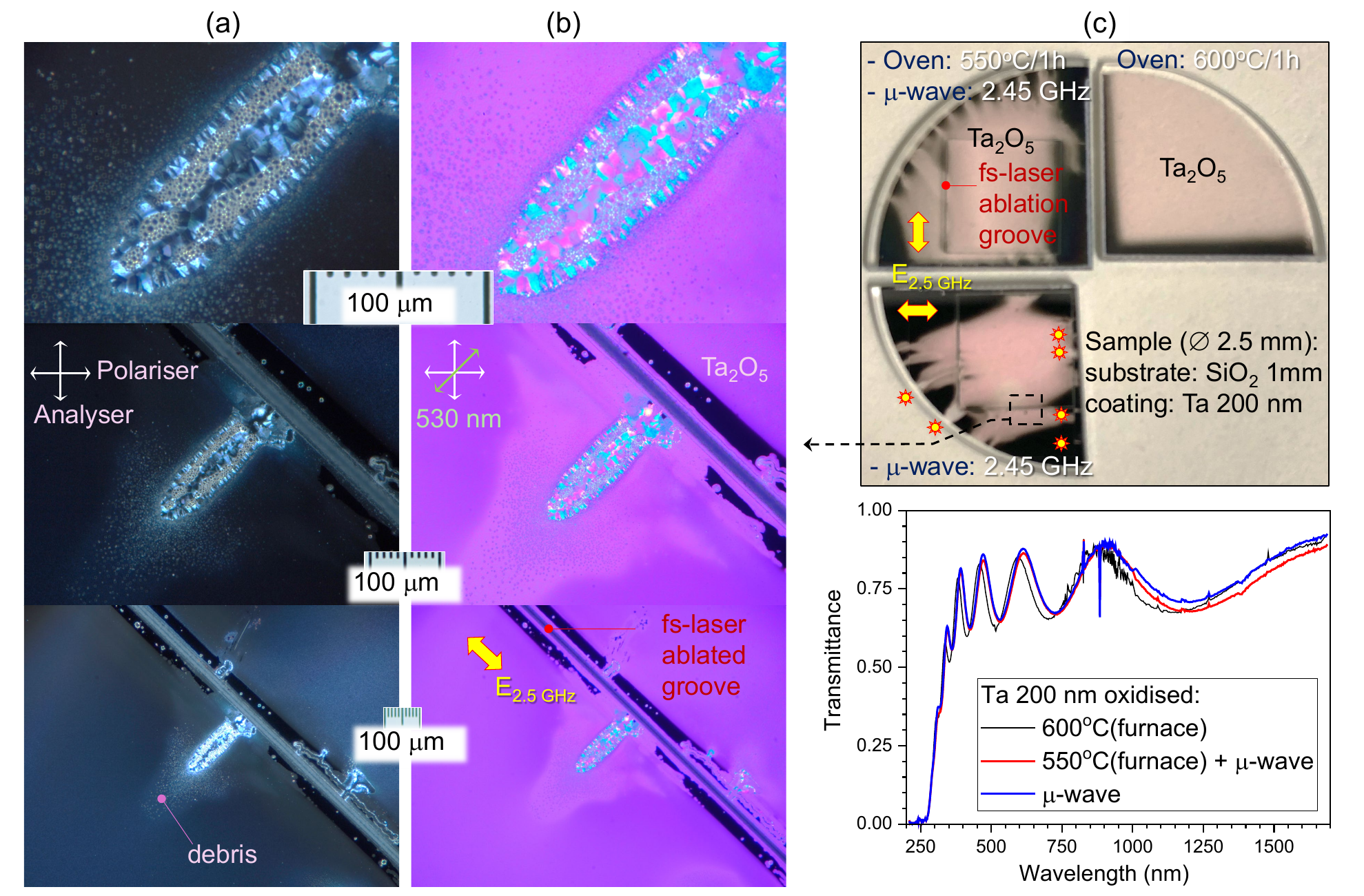}
\caption{\label{f-ox} Oxidation of a 200-nm-thick Ta film on a \ce{SiO2} substrate. (a) Cross-polarized optical image of a location where explosive microwave energy deposition occurred. (b) Cross-polarized image with a $\lambda$-plate (530~nm), revealing regions of increased and decreased optical retardance. (c) Photograph of the sample quarters subjected to different oxidation treatments: (1) furnace heating at 600$^\circ$C/1~h for full oxidation; (2) 550$^\circ$C/1~h (no oxidation) followed by microwave exposure in steps up to $\sim 50$~W; and (3) microwave exposure alone for full oxidation. Star markers indicate locations of strong energy deposition. The bottom inset shows the transmittance of the three quarters oxidized by the different protocols; the ellipsometer was operated in transmission mode for p-polarization.}
\end{figure*}

Figure~\ref{f-ox} shows the appearance of a 200~nm Ta film oxidized by exposure in the microwave cavity, reached through several gradual annealing steps of increasing power up to a final 50~W. A separate experiment in a standard tube furnace under \ce{O2} flow confirmed that 1~hour at $600^\circ$C is required for the complete conversion of Ta into \ce{Ta2O5}; at $550^\circ$C the transmittance was unchanged across the visible range and only slightly increased in the near-IR~\cite{26olt115018}. In the microwave cavity, by contrast, oxidation of extended areas took only tens of seconds, although the resulting oxide film was not uniform.

The pattern of non-oxidized (black) regions in Fig.~\ref{f-ox}(c) has a characteristic structure set by the cavity mode (see the finite element method (FEM) simulations and the discussion in Sec.~\ref{disco}). The central cavity mode inside the \ce{SiO2} tube has a horizontal electric $E$-field. Once an object is inserted into the cavity, the local $E$-field distribution is redistributed according to the geometry of the sample and its complex permittivity $\varepsilon^* = \varepsilon' + i\varepsilon''$.

\begin{figure*}[tb]
\centering\includegraphics[width=1\textwidth]{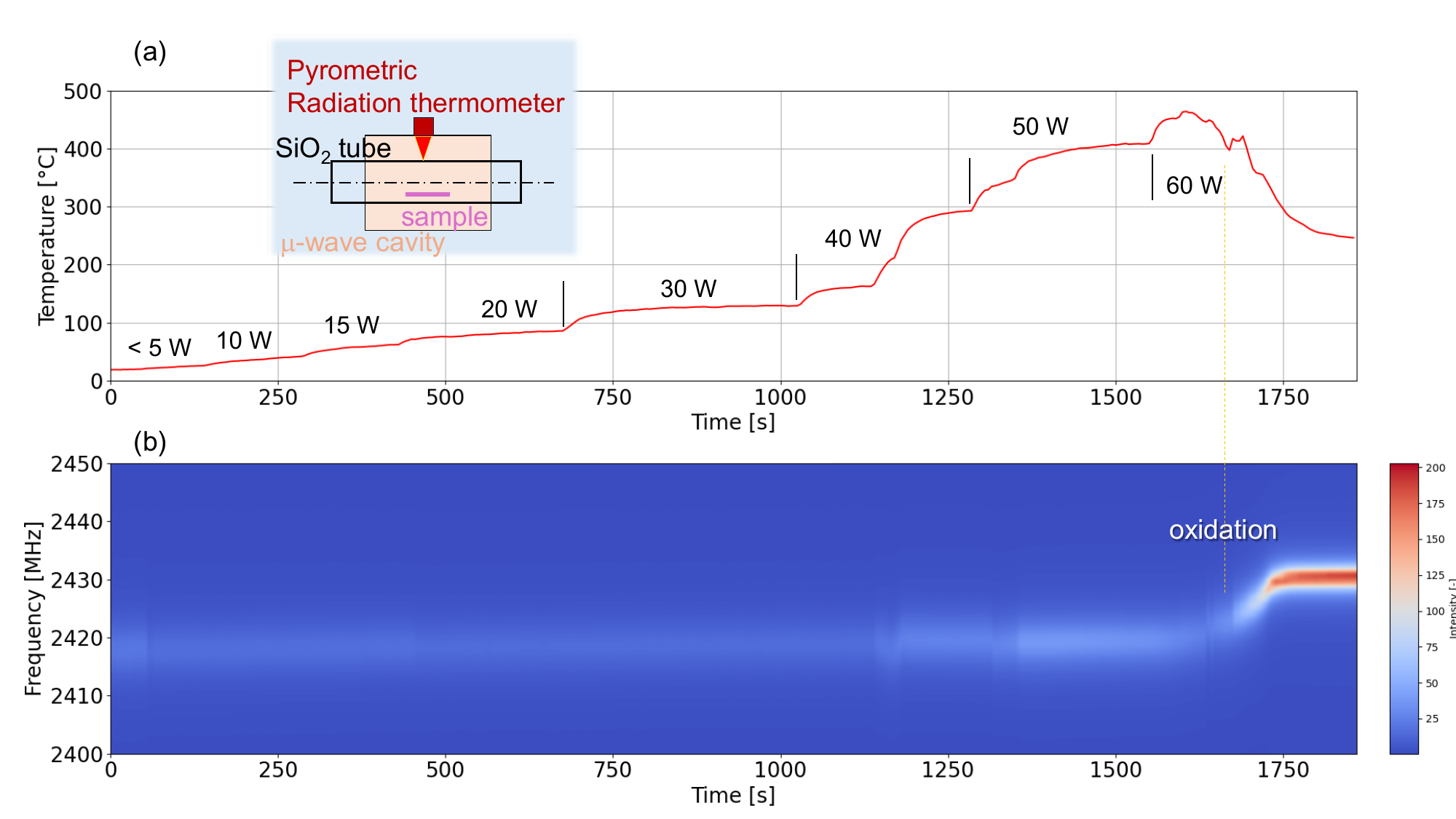}
\caption{\label{f-freq} Data log of temperature and cavity resonance during microwave-induced oxidation of a 200~nm Ta film on \ce{SiO2}. (a) Temperature evolution measured at the outer wall of the \ce{SiO2} tube (see inset) as the microwave power was raised in several steps to a final 60~W. (b) Corresponding evolution of the cavity resonance frequency and amplitude. Sample: previously annealed at 550$^\circ$C (see Fig.~\ref{f-ox}(c)).}\end{figure*}

The sites at which strong oxidation was initiated are surrounded by ablated regions and debris fields (Fig.~\ref{f-ox}(a,b)). In a film as thin as 200~nm, the exothermic oxidation behaves like a nano-thermite~\cite{21t34} and partially removes the film. The cross-polarized image recorded with a 530~nm waveplate shows colors set by the local optical path $nd$, where $n$ is the refractive index and $d$ the thickness; any birefringence $\Delta n$ present shifts the local index to $n_0\pm\Delta n$.

\begin{figure*}[tb]
\centering\includegraphics[width=0.9\textwidth]{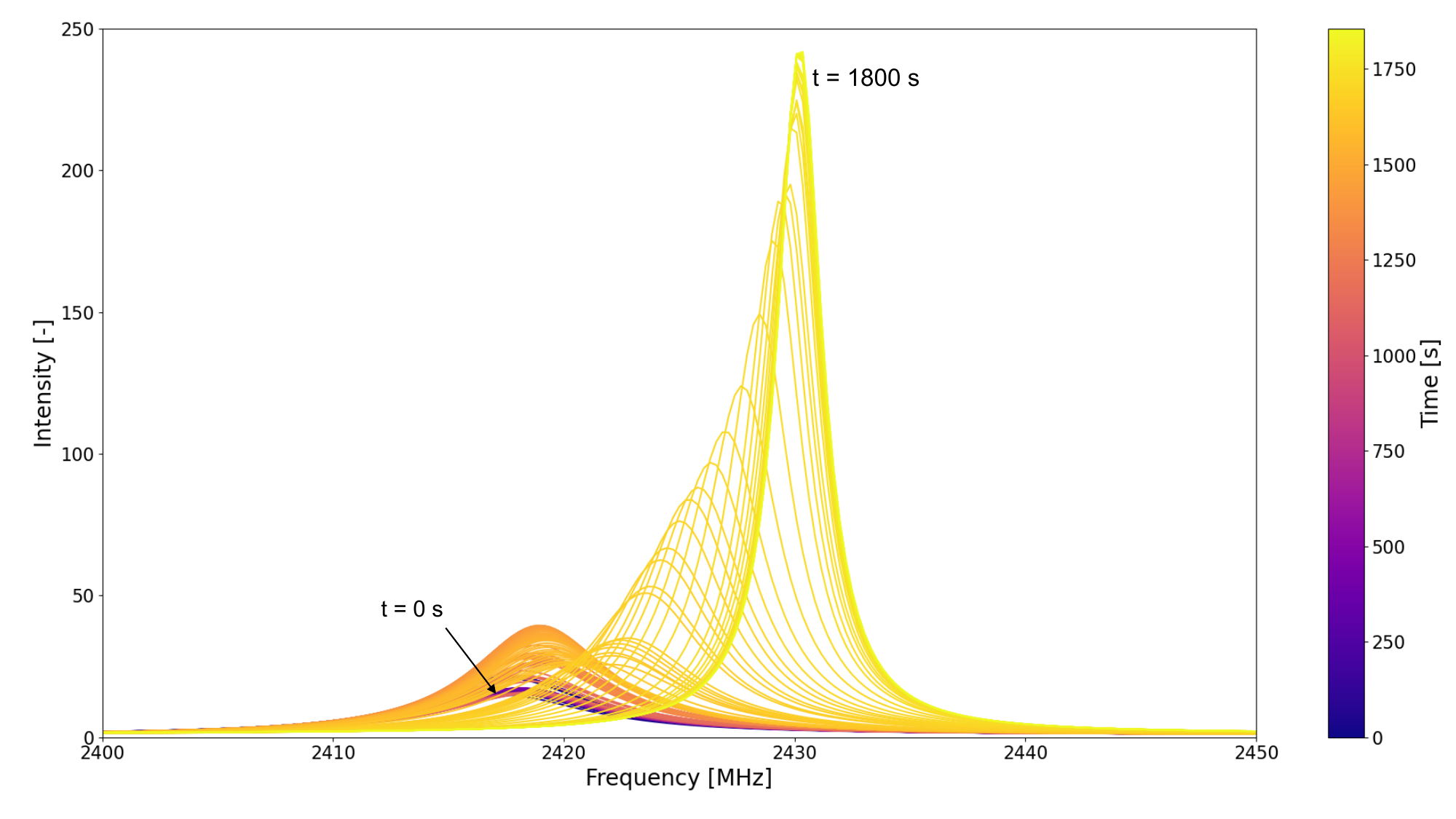}
\caption{\label{f-reso} Temporal evolution of the cavity mode during the increase of power and temperature and the oxidation of Ta (same run as in Fig.~\ref{f-freq}(b)). Sample: previously annealed at 550$^\circ$C (see Fig.~\ref{f-ox}(c)).}\end{figure*}

\begin{figure*}[tb]
\centering\includegraphics[width=1\textwidth]{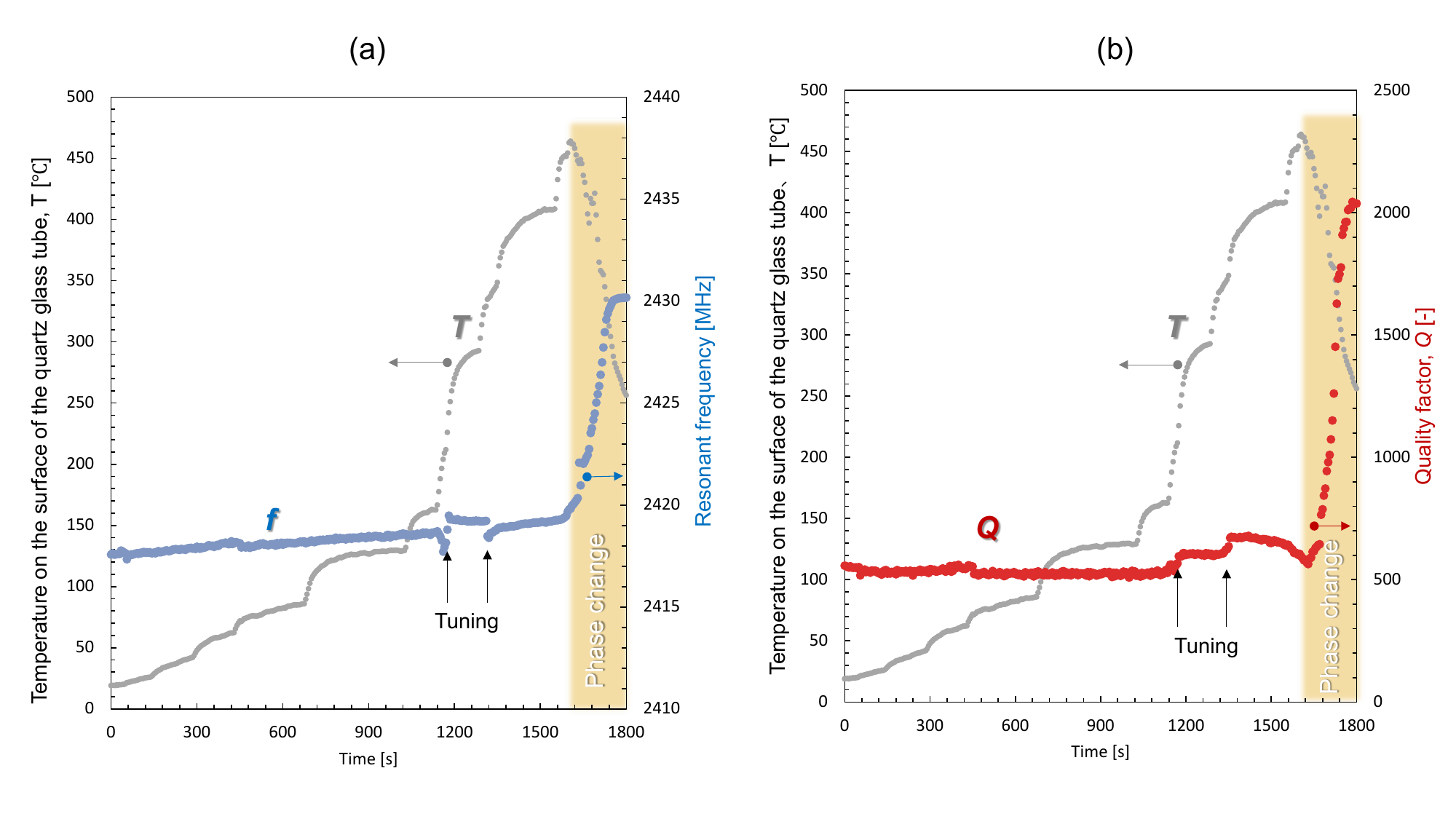}
\caption{\label{f-sum} Summary of the oxidation of a 200~nm Ta film: (a) temperature and resonant frequency, and (b) temperature and 
quality factor $Q$, 
versus microwave exposure time at increasing powers up to 60~W. The markers of the tuning moments correspond to impedance-matching adjustments made to reduce the fraction of energy reflected 
back from the cavity. 
The strongest changes in $f$ and $Q$, occurring once the maximum temperature is reached, mark the metal-to-oxide transformation (shaded region).}\end{figure*}

Figure~\ref{f-freq} shows the measured evolution of temperature and resonant frequency as the microwave power was increased in steps. The pyrometer read the radiative temperature on the outer surface of the \ce{SiO2} tube directly above the sample and was calibrated to the emissivity of \ce{SiO2} over the 2-to-6.8~$\mu$m window. The highest temperature reached was $\sim 450^\circ$C, measured approximately 1~cm above the surface of the Ta sample. Figure~\ref{f-freq}(b) and Fig.~\ref{f-reso} show the resonance evolving from broad and low in amplitude at the onset of microwave heating to narrow and high in amplitude during and after oxidation. The width of the resonance, and hence the quality factor $Q$, is governed by the imaginary part of the permittivity $\varepsilon''$. The combined evolution of temperature, resonant frequency, and $Q$-factor 
during heating and oxidation is summarized in Fig.~\ref{f-sum}. The resonant frequency $f_r$ rises gradually with temperature and then jumps by 10~MHz at the onset of oxidation. 

The microwave response recorded during oxidation of the Ta nano-film reveals a clear correlation between the chemical transformation of the film and the electromagnetic behavior of the resonant cavity. As heating proceeds, both the resonance frequency and the cavity $Q$-factor increase progressively, indicating a simultaneous reduction of cavity loading and of microwave dissipation (Fig.~\ref{f-sum}).

The rise in resonance frequency indicates that the electromagnetic perturbation introduced by the sample weakens as oxidation proceeds. Initially, the metallic Ta film interacts strongly with the electric field of the TM$_{010}$ cavity mode because of its high electrical conductivity. In this state the film acts as a highly dissipative electromagnetic load, 
supporting induced currents and partially screening the electric field within the cavity. Both effects increase the effective electromagnetic loading of the resonator and push the resonance frequency down.

As oxidation progresses, metallic Ta is gradually converted into \ce{Ta2O5}, whose electrical conductivity and microwave losses are far lower. The reduced free-carrier density suppresses the induced conductive currents in the film and thereby weakens the perturbation of the cavity, so the resonance shifts toward higher frequency as the cavity approaches a less-loaded state.

The cavity $Q$-factor increases in parallel throughout the oxidation. Since $Q$ is inversely related to the dissipative losses, this behavior signals a progressive reduction of microwave absorption in the sample. The metallic Ta layer initially introduces strong ohmic losses through the surface currents induced by the electric field parallel to the film. As oxidation disrupts the conductive metallic network, these ohmic dissipation channels are progressively suppressed and the cavity $Q$-factor rises substantially.

The observed behavior is therefore consistent with a gradual transition from a conductive metallic state to a dielectric oxide state. The simultaneous increase of resonance frequency and $Q$-factor (Fig.~\ref{f-sum}) points to the reduction of electrical conductivity and microwave loss during oxide formation as the dominant mechanism.

\section{Discussion}\label{disco}

\subsection{Annealing and oxidation in the $\mu$-wave cavity}
\begin{figure*}[tb]
\centering\includegraphics[width=1\textwidth]{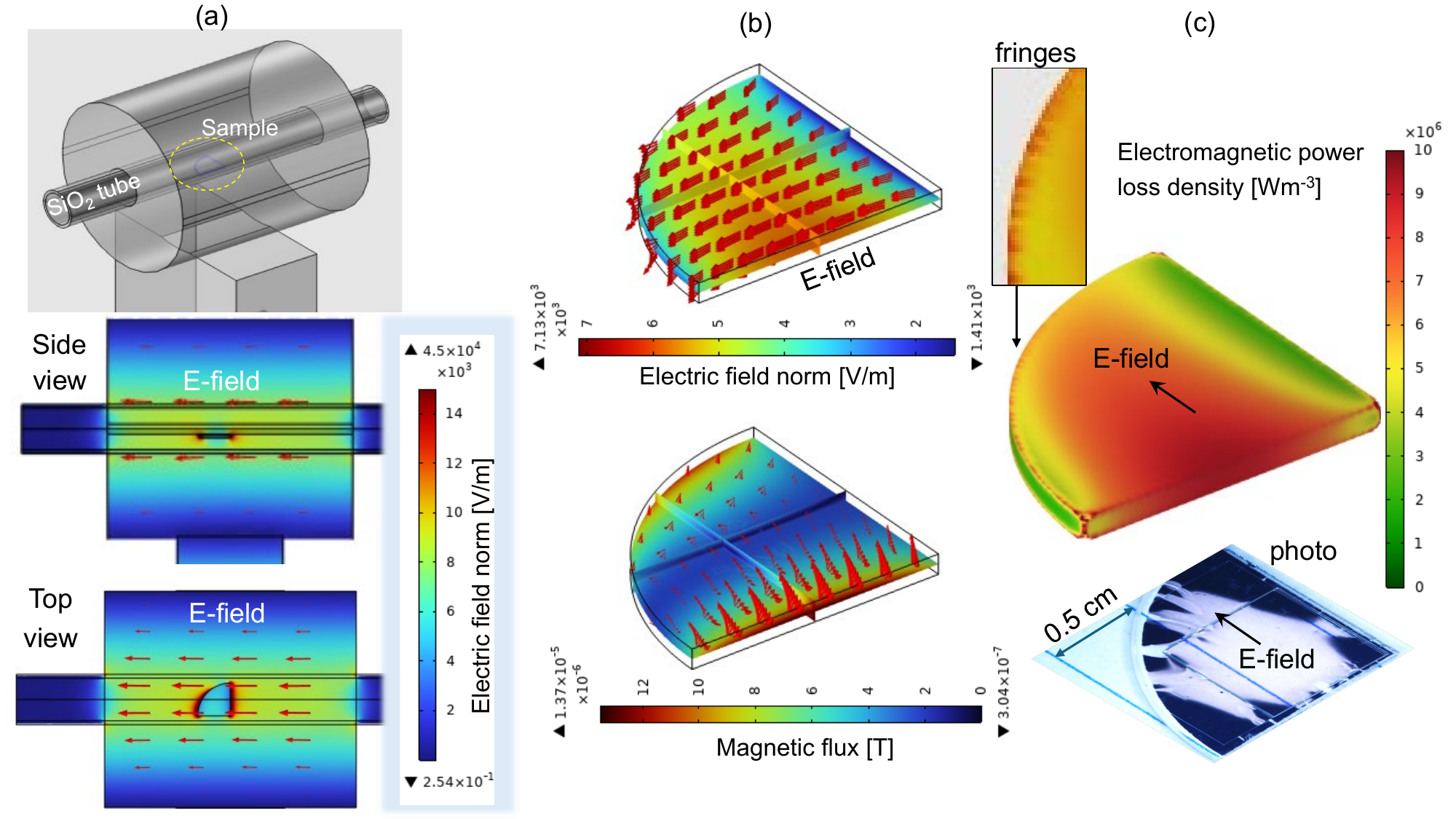}
\caption{\label{f-sim} FEM simulations of the 2.45~GHz resonator/cavity (COMSOL Multiphysics 6.3). (a) Geometry and E-field distributions; the input power at the coaxial port is 1~W. (b) E-field strength [V/m] and magnetic flux [T] for the sample used. 
(c) Energy deposition presented as electromagnetic power loss density [W/m$^3$], together with a photograph of the actual sample oxidized in the microwave cavity. The zoomed-in edge of the sample shows a fringe pattern that was independent of the FEM meshing.}\end{figure*}

Figure~\ref{f-sim} shows the FEM results for the sample used here: a 200~nm Ta film on a 1-mm-thick \ce{SiO2} substrate placed in the 2.45~GHz cavity. The permittivity of the sample determined by VNA was used as the input for the simulations.
The predicted locations of energy deposition (power loss density) qualitatively match the modifications observed on the actual sample. The E-field is concentrated on the axis of the cylindrical cavity, where the sample is inserted, and the edges of the sample lying along the horizontal E-field show pronounced field localization and enhancement.

The volumetric power absorption density of microwave energy in a dielectric material is governed by the interaction between the applied electric field and the dielectric properties of the material, 
\begin{equation}\label{e-pabs}
P = \tfrac{1}{2}\,\omega\,\varepsilon_0\,\varepsilon''\,|E|^2,
\end{equation}
where $\varepsilon_0 = 8.854\times 10^{-12}~\mathrm{F/m}$ is the permittivity of free space, $\varepsilon''$ is the dielectric loss factor, 
and $|E|$ is the amplitude of the electric field inside the medium, consistent with Eq.~\ref{e-h}. The absorbed microwave power ultimately appears as volumetric heat generation. Equating $P$ with the rate at which thermal energy is stored in the material gives
\begin{equation}
P = \rho C_p \frac{dT}{dt},
\end{equation}
where $\rho$ $[\mathrm{kg/m^3}]$ is the material density, $C_p$ $[\mathrm{J/(kg \cdot K)}]$ is the specific heat capacity, and $dT/dt$ $[\mathrm{K/s}]$ is the local rate of temperature rise. This relation links the electromagnetic and thermal problems directly in the analysis of microwave heating.

Signatures of surface breakdown along the circular edge of the sample (Fig.~\ref{f-ox}(c)) resemble the pattern predicted by the model (Fig.~\ref{f-sim}(c)). Runs with different meshes confirmed that the alternating micro-regions of energy deposition at the edge are a physical feature and not a meshing artifact. The agreement between simulation and experiment remains qualitative, however, because the material properties change continuously during heating and oxidation and this evolution was not fed back into the model.

\subsection{Perspective of 3D localization of microwave HTA}
Fast HTA by microwave exposure could be tested for the controlled precipitation of Ca, Ba, and Li colloidal nanoparticles in regions where fs-laser-inscribed nanogratings have been formed inside the corresponding fluorides~\cite{kyoto}, materials that find applications over the very wide 0.15-15~$\mu$m spectral range, from UV-C to the mid-IR. Localized heating, structural restoration, and defect annealing in regions damaged by fs-laser patterning are a promising direction to explore for controlling the UV-visible emission of F and F$_2$ color centers. The emission of fluorides doped with praseodymium Pr$^{3+}$ rare-earth ions, which are cascade emitters under vacuum-UV excitation~\cite{KUCK}, could likewise be defined in 3D by laser inscription followed by localized microwave HTA.

All three pillars of quantum optics --- collective emission, single-ion emission, and ensemble-based quantum memories~\cite{qmem} --- would benefit from spatially controlled, laser-inscribed functional modifications combined with microwave HTA, a technique uniquely suited to nanoscale localized energy deposition and to the control of thermally driven modifications of materials. Rare-earth-doped fluorides are increasingly used in quantum optics because the shielded 4f orbitals of lanthanide ions confer long-lived radiative transitions, narrow homogeneous linewidths, and rich hyperfine structures that underpin long optical and spin coherence~\cite{lanth}. Since lanthanides are heavy, with atomic masses $> 139$, compared with the host atoms in fluorides, where fluorine has a relative atomic mass of $\sim 19$, fs-laser 3D inscription can be used for efficient mass separation~\cite{11nc445} and hence for hyper-doping. Microwave HTA could also prove effective in restoring the crystal lattice after ion implantation, which is used for nanoscale precision in ion deposition and doping as well as for thermal control of nanoparticle growth and assembly~\cite{24as2402840}. Fs-laser post-processing after Au implantation adds control over optical anisotropy and nonlinearity, 
which has been used to realize a Q-switched laser in a plasmonic-crystalline matrix~\cite{25s2411607}. Such a complex, optically active matrix was formed by nanoscale-localized volumetric energy deposition under fs-laser 3D-controlled irradiation, and could be complemented by microwave-HTA-guided modification of the laser-inscribed patterns.

\subsection{Two-step oxide patterning}
A concept of future interest is a \emph{two-step} route: a laser modifies, or gently sub-ablatively damages, a metal surface in defined regions, and a \emph{separate} thermal anneal in an oxidizing atmosphere then grows oxide preferentially in those regions. The furnace-based version of this route has already been demonstrated, as summarized below; replacing the furnace step with microwave annealing is the natural extension.
Laser chemical machining of \mbox{42CrMo4} steel provides one example, in which the laser-modified rim zone --- with altered residual stress, roughness, and seed oxide --- acts as a set of nucleation sites that thicken the oxide grown during a subsequent oxidation at $500\,^{\circ}$C in air~\cite{schupp2021oxidation}. A clean ``the laser defines the geometry, the furnace does the chemistry'' example is direct laser interference patterning (DLIP) of thin tantalum films, which are afterwards converted into transparent oxide gratings by thermal-oxidation annealing~\cite{nikitina2024submicron}.
Closely related are single-step results in which the laser itself drives the oxidation with no separate furnace step: selective laser oxidation of thin chromium films followed by chemical development~\cite{gedvilas2017thermochemical}, and laser oxidation of titanium producing interference-colored oxide layers~\cite{perezdelpino2002coloring}.
These observations rest on well-understood kinetics, including the way laser-injected impurities and defects accelerate diffusion and oxide growth relative to isothermal oxidation~\cite{nanai1997laser}. Laser-induced defect engineering of an oxide, illustrating how laser-created defects tune the subsequent oxidation and catalytic reactivity, has been shown for cobalt oxide platelets~\cite{schellenburg2026mechanistic}.

\section{Conclusions and Outlook}

Microwave energy deposition can form a Ta oxide nano-film within tens of seconds at an applied power of $\sim 50$~W. By measuring the resonance frequency and the $Q$-factor 
it was possible to track 
the changes and thus to follow the state of the material in real time. The changes in 
$f$ and $Q$ extracted from the line shape and spectral position of the cavity resonance are directly related to the volume, the state (solid or liquid), mass loss through evaporation or mass gain through oxidation or nitridation, and emissivity; that is, to material properties encoded in the permittivity. These quantities are interlinked, and together they provide a feedback channel for monitoring changes during HTA of complex samples in real time. Modifications of optical properties that are usually achieved by conventional furnace HTA of an entire sample or micro-optical device could therefore also be made by microwave heating with highly localized energy deposition through dipole or Joule mechanisms. Nano- and micro-sized objects, with their large surface-to-volume ratio, are particularly promising candidates for microwave annealing.

Overall, these results demonstrate that cavity perturbation measurements performed under microwave irradiation can sensitively track dynamic physicochemical transformations in thin films. The simultaneous evolution of the resonance frequency and the $Q$-factor gives direct insight into the coupling between oxidation kinetics, electrical conductivity, and microwave electromagnetic response during high-temperature processing.

\begin{acknowledgments}
\small
M.Z. acknowledges support via  MEXT Next-Generation Computational Science Grand Reach Program.  L.G. received funding from the Research Council of Lithuania (LMTLT), agreement No. S-PD-24-94. S.J. acknowledges support via ARC grant DP240103231. H.-H.H. is grateful for a research stay at the Laser Research Center, Vilnius University, in 2025. J.M. acknowledges support via JST CREST (Grant No. JPMJCR19I3) and KAKENHI (No. 26K01228).
\end{acknowledgments}

\bibliography{aipsamp}

@PREAMBLE{
 "\providecommand{\noopsort}[1]{}" 
 # "\providecommand{\singleletter}[1]{#1}%" 
}

@article{Massi,
title = {Microwave-assisted dehydration of calcium hydroxide for thermochemical energy storage},
journal = {Journal of Energy Storage},
volume = {108},
pages = {115195},
year = {2025},
issn = {2352-152X},
doi = {https://doi.org/10.1016/j.est.2024.115195},
url = {https://www.sciencedirect.com/science/article/pii/S2352152X24047819},
author = {Massimiliano Zamengo and Hisahiro Einaga and Yuji Wada and Junko Morikawa}
}

@article{25s2411607,
  title = {Strong {{Polarization}}-{{Tuned Optical Nonlinearity Via Femtosecond}}-{{Laser Plasmonic Nanolithography}} in {{Lithium Niobate}}},

  author = {Zhu, Han and Sun, Wenqing and Chu, Lingrui and Zhou, Shengqiang and Wu, Tianci and Ye, Qingchuan and Sun, Xiaoli and Juodkazis, Saulius and Chen, Feng},
  year = {2025},
  month = jun,
  journal = {Small},
  volume = {21},
  number = {25},
  pages = {2411607},
doi = {https://doi.org/10.1002/smll.202411607}
}

@article{toyota,
title = {Microwave Processing and Applications to Future Automobile},
journal = {R\&D Rev. Toyota CRDL},
volume = {43}, issue = {3},
pages = {75 - 92},
year = {2015},
issn = {0017-9310},
doi = {https://www.tytlabs.co.jp; visited April 1, 2026},
author = {H Fukushima}
}

@Article{Hamashima,
author ="Hamashima, Tatsuya and Nishioka, Masateru and Sugiyama, Takeharu and Watanabe, Ken and Hojo, Hajime and Einaga, Hisahiro",
title  ="Understanding the microwave heating properties of La–Ce–Ni oxides based on structural{,} dielectric{,} and conductive analysis",
journal  ="Phys. Chem. Chem. Phys.",
year  ="2025",
volume  ="27",
issue  ="44",
pages  ="23802-23812",
publisher  ="The Royal Society of Chemistry",
doi  ="10.1039/D5CP02279G",
url  ="http://dx.doi.org/10.1039/D5CP02279G"}

@article{sic,
    author = {Sundaresan, Siddarth G. and Rao, Mulpuri V. and Tian, Yong-lai and Ridgway, Mark C. and Schreifels, John A. and Kopanski, Joseph J.},
    title = {Ultrahigh-temperature microwave annealing of {Al+}- and {P+}-implanted 4H-SiC},
    journal = {Journal of Applied Physics},
    volume = {101},
    number = {7},
    pages = {073708},
    year = {2007}
}

@article{pero,
title = {Rapid microwave annealing of perovskite films: Exploring the mechanism of heat generation and influence on growth kinetics},
journal = {Solar Energy Materials and Solar Cells},
volume = {295},
pages = {113967},
year = {2026},
issn = {0927-0248},
doi = {https://doi.org/10.1016/j.solmat.2025.113967},
url = {https://www.sciencedirect.com/science/article/pii/S0927024825005689},
author = {Syed Nazmus Sakib and David N.R. Payne and Jincheol Kim and Shujuan Huang and Binesh Puthen Veettil}
}

@article{polo,
author = {Truong, Thien and Liang, Wensheng and Basnet, Rabin and Nemeth, William and Stradins, Pauls and Young, David L. and Macdonald, Daniel and Fong, Kean Chern},
title = {Microwave Annealing for Fast and Effective Hydrogen Activation in Polycrystalline Silicon Passivating Contacts},
journal = {Advanced Energy and Sustainability Research},
volume = {6},
number = {10},
pages = {2500004},
year = {2025}
}

@Article{park,
AUTHOR = {Park, Ki-Woong and Cho, Won-Ju},
TITLE = {Thermal Damage-Free Microwave Annealing with Efficient Energy Conversion for Fabricating of High-Performance a-IGZO Thin-Film Transistors on Flexible Substrates},
JOURNAL = {Materials},
VOLUME = {14},
YEAR = {2021},
NUMBER = {10},
ARTICLE-NUMBER = {2630},
URL = {https://www.mdpi.com/1996-1944/14/10/2630},
PubMedID = {34069832},
ISSN = {1996-1944},
DOI = {10.3390/ma14102630}
}

@article{26olt115018,
title = {Surface texturing and localized oxidation of Tantalum nano-film by fs-laser},
journal = {Optics \& Laser Technology},
volume = {199},
pages = {115018},
year = {2026},
author = {Lina Grineviciute and Hsin-Hui Huang and Haoran Mu and Julianija Nikitina and Nguyen Hoai An Le and Tomas Katkus and Andrew Siao Ming Ang and Saulius Juodkazis}
}

@article{jung,
author = {Jung, Minhyun and Kim, Chaeheon and Hwang, Junghyeon and Kim, Giuk and Shin, Hunbeom and Gaddam, Venkateswarlu and Jeon, Sanghun},
title = {High Pressure Microwave Annealing Effect on Electrical Properties of {HfxZr1–xO} Films near Morphotropic Phase Boundary},
journal = {ACS Applied Electronic Materials},
volume = {5},
number = {9},
pages = {4826-4835},
year = {2023},
doi = {10.1021/acsaelm.3c00623},

URL = { 
    
        https://doi.org/10.1021/acsaelm.3c00623
    
    

},
eprint = { 
    
        https://doi.org/10.1021/acsaelm.3c00623
    
    

}

}

@article{solvo,
author = {Hwang, Jongkook and Chun, Jinyoung},
title = {Microwave-assisted solvothermal synthesis of sodium metal fluoride (NaxMFy) nanopowders},
journal = {Journal of the American Ceramic Society},
volume = {102},
number = {11},
pages = {6475-6479},
year = {2019}
}

@article{rare,
title = {Influence of microwave heating on the extractions of fluorine and Rare Earth elements from mixed rare earth concentrate},
journal = {Hydrometallurgy},
volume = {162},
pages = {104-110},
year = {2016},
issn = {0304-386X},
doi = {https://doi.org/10.1016/j.hydromet.2016.03.022},
url = {https://www.sciencedirect.com/science/article/pii/S0304386X16301177},
author = {Yukun Huang and Ting-an Zhang and Zhihe Dou and Jiang Liu and Junhang Tian}
}

@article{defl,
title = {Defluorination study of spent carbon cathode by microwave high-temperature roasting},
journal = {Journal of Environmental Management},
volume = {302},
pages = {114028},
year = {2022},
issn = {0301-4797},
doi = {https://doi.org/10.1016/j.jenvman.2021.114028},
url = {https://www.sciencedirect.com/science/article/pii/S0301479721020909},
author = {Zhi Zhu and Lei Xu and Zhaohui Han and Jianhua Liu and Libo Zhang and Chuxuan Yang and Zhangbiao Xu and Peng Liu}
}

@article{kyoto,
author = {Y Toyama and Y Shimotsuma and H Kitahara and M Shimizu and K Miura  and M Tani},
title = {Photoinduced Structural Changes in Fluoride Single Crystals},
journal = {Journal of Laser Micro/Nanoengineering},
volume = {21},
number = {2},
pages = {accepted},
year = {2026}
}

@article{KUCK,
title = {Photon cascade emission in Pr3+-doped fluorides},
journal = {Journal of Luminescence},
volume = {102-103},
pages = {176-181},
year = {2003},
note = {Proceedings of the 2002 International Conference on Luminescence and Optical Spectroscopy of Condensed Matter},
issn = {0022-2313},
doi = {https://doi.org/10.1016/S0022-2313(02)00486-6},
url = {https://www.sciencedirect.com/science/article/pii/S0022231302004866},
author = {S. Kück and I. Sokólska and M. Henke and M. Döring and T. Scheffler}
}

@article{lanth,
author = {Huang, Kai and Fung-A-Fat, Joshua and Wu, Jiaze and Yu, Shupei and Fung-A-Fat, Daniel and Deng, Katherine and Zuercher, Gwenyth and Sankar, Diya and Saroha, Ishana and Xu, Weichu and Han, Gang},
title = {Lanthanide-Based Quantum Optical Materials},
journal = {Advanced Functional Materials},
volume = {36},
number = {23},
pages = {e24562},
doi = {https://doi.org/10.1002/adfm.202524562},
url = {https://advanced.onlinelibrary.wiley.com/doi/abs/10.1002/adfm.202524562},
eprint = {https://advanced.onlinelibrary.wiley.com/doi/pdf/10.1002/adfm.202524562},
year = {2026}
}

@inbook{qmem,
author = {Jing, Bo and Bao, Xiao-Hui},
publisher = {John Wiley \& Sons, Ltd},
isbn = {9783527837427},
title = {Ensemble-Based Quantum Memory: Principle, Advance, and Application},
booktitle = {Photonic Quantum Technologies},
chapter = {17},
pages = {433-462},
year = {2023}
}

@article{11nc445,
author = {A Vailionis and EG Gamaly and V Mizeikis and W Yang and AV Rode and S Juodkazis },
title = {Evidence of superdense aluminium synthesized by ultrafast microexplosion},
journal = {Nature Communications },
volume = {2},
pages = {445},
year = {2011}
}

@article{24as2402840,
author = {Zhu, Han and Chu, Lingrui and Lv, Hengyue and Ye, Qingchuan and Juodkazis, Saulius and Chen, Feng},
title = {Ultrafast Laser Manipulation of In-Lattice Plasmonic Nanoparticles},
journal = {Advanced Science},
volume = {11},
number = {38},
pages = {2402840},
year = {2024}
}

@article{mw,
author = {Truscott, Benjamin S. and Kelly, Mark W. and Potter, Katie J. and Johnson, Mack and Ashfold, Michael N. R. and Mankelevich, Yuri A.},
title = {Microwave Plasma-Activated Chemical Vapor Deposition of Nitrogen-Doped Diamond. I. \ce{N2}/\ce{H2} and \ce{NH3}/\ce{H2} Plasmas},
journal = {Journal of Physical Chemistry A},
volume = {119},
number = {52},
pages = {12962-12976},
year = {2015}
}

@misc{26arX,
      title={Burst-mode fs-laser direct writing for full-thickness oxidation of Ta thin films}, 
      author={Lina Grineviciute and Hsin-Hui Huang and Haoran Mu and Nguyen Hoai An Le and Andrew Siao Ming Ang and Dan Kapsaskis and Tomas Katkus and Saulius Juodkazis},
      year={2026},
      eprint={2602.06444},
      archivePrefix={arXiv},
      primaryClass={physics.optics},
      url={https://arxiv.org/abs/2602.06444}, 
}

@Article{21t34,
AUTHOR = {Lundgaard, Stefan and Ng, Soon Hock and Cahill, Damien and Dahlberg, Johan and Allender, Jamie and Barber, Michael and Stephens, Joshua and Juodkazis, Saulius},
TITLE = {Electrical Breakdown Spectroscopy of Nano-/Micro-Thermites},
JOURNAL = {Technologies},
VOLUME = {9},
YEAR = {2021},
NUMBER = {2},
pages = {34}
}

@article{sicA,
author = {Rabby, Md. Reza-E- and Jeelani, Shaik and Rangari, Vijaya K.},
title = {Microwave processing of SiC nanoparticles infused polymer composites: Comparison of thermal and mechanical properties},
journal = {Journal of Applied Polymer Science},
volume = {132},
number = {12},
pages = {},
year = {2015}
}

@article{Wang2026,
  author  = {Wang, Y. and others},
  title   = {Microwave synthesis of gram-scale, millimeter-size layered transition metal oxide crystals},
  journal = {NPG Asia Materials},
  volume  = {18},
  pages   = {8},
  year    = {2026},
  doi     = {10.1038/s41427-026-00637-8}
}

@article{Ding2025,
  author  = {Ding, Shu Ming and Wang, Xue Nan and Shao, Yan and Wang, Bo and Wang, Zhan Jie},
  title   = {Effect of microwave annealing on the polarization properties of sol-gel derived {Pb(Zr$_{0.4}$Ti$_{0.6}$)O$_3$} epitaxial thin films},
  journal = {Materials Today Communications},
  volume  = {46},
  pages   = {112518},
  year    = {2025},
  doi     = {10.1016/j.mtcomm.2025.112518}
}

@article{Cui2023,
  author  = {Cui, Siwei and Yang, Hui and Zhang, Yifei and Su, Xing and Wu, Dongping},
  title   = {Effect of microwave annealing on the sensing characteristics of {HfO$_2$} thin film for high sensitive {pH-EGFET} sensor},
  journal = {Micromachines},
  volume  = {14},
  number  = {10},
  pages   = {1854},
  year    = {2023},
  doi     = {10.3390/mi14101854}
}

@article{Hwang2024,
  author  = {Hwang, Chang-Hyun and Beak, Jung-Hoon and Kim, Sung-Il and Kim, Soo-Young},
  title   = {Effect of pyrolysis temperature on microwave heating properties of oxidation-cured polycarbosilane powder},
  journal = {Crystals},
  volume  = {14},
  number  = {12},
  pages   = {1080},
  year    = {2024},
  doi     = {10.3390/cryst14121080}
}

@article{McFadzean2025,
  author  = {McFadzean, Ross George Bell and Smith, Ronald and Drysdale, Timothy David and Gregory, Duncan H.},
  title   = {A reactor for in situ, time-resolved neutron diffraction studies of microwave-induced rapid solid-state chemical reactions},
  journal = {Philosophical Transactions of the Royal Society A},
  volume  = {383},
  number  = {2297},
  year    = {2025},
  doi     = {10.1098/rsta.2024.0065}
}

@article{schupp2021oxidation,
  author  = {Schupp, Alexander and P{\"u}tz, Ren{\'e} Daniel and Beyss, Oliver
             and Beste, Lucas-Hermann and Radel, Tim and Zander, Daniela},
  title   = {Change of Oxidation Mechanisms by Laser Chemical Machined Rim
             Zone Modifications of {42CrMo4} Steel},
  journal = {Materials},
  year    = {2021},
  volume  = {14},
  number  = {14},
  pages   = {3910},
  doi     = {10.3390/ma14143910}
}

@article{nikitina2024submicron,
  author  = {Nikitina, Julianija and Indri{\v s}i{\=u}nas, Simonas
             and Tolenis, Tomas and Andrulevi{\v c}ius, Mindaugas
             and Grinevi{\v c}i{\=u}t{\.e}, Lina},
  title   = {Submicron Periodic Structures in Metal Oxide Coating via Laser
             Ablation and Thermal Oxidation},
  journal = {Applied Surface Science Advances},
  year    = {2024},
  volume  = {24},
  pages   = {100660},
  doi     = {10.1016/j.apsadv.2024.100660}
}

@article{gedvilas2017thermochemical,
  author  = {Gedvilas, Mindaugas and Voisiat, Bogdan and Indri{\v s}i{\=u}nas, Simonas
             and Ra{\v c}iukaitis, Gediminas and Veiko, Vadim and Zakoldaev, Roman
             and Sinev, Dmitry and Shakhno, Elena},
  title   = {Thermo-Chemical Microstructuring of Thin Metal Films Using
             Multi-Beam Interference by Short (Nano- \& Picosecond) Laser Pulses},
  journal = {Thin Solid Films},
  year    = {2017},
  volume  = {634},
  pages   = {134--140},
  note    = {DOI via Elsevier PII S0040609017303437}
}

@article{perezdelpino2002coloring,
  author  = {P{\'e}rez del Pino, A. and Serra, P. and Morenza, J. L.},
  title   = {Coloring of Titanium by Pulsed Laser Processing in Air},
  journal = {Thin Solid Films},
  year    = {2002},
  volume  = {415},
  number  = {1--2},
  pages   = {201--205},
  doi     = {10.1016/S0040-6090(02)00632-6}
}

@article{nanai1997laser,
  author  = {N{\'a}nai, L. and Vajtai, R. and George, T. F.},
  title   = {Laser-Induced Oxidation of Metals: State of the Art},
  journal = {Thin Solid Films},
  year    = {1997},
  volume  = {298},
  number  = {1--2},
  pages   = {160--164},
  doi     = {10.1016/S0040-6090(96)09390-X}
}

@article{schellenburg2026mechanistic,
  author  = {Schellenburg and others},
  title   = {Mechanistic Understanding of Laser-Induced Defect Engineering of
             Anisotropic Cobalt Oxide Spinel Platelets in Water},
  journal = {ChemCatChem},
  year    = {2026},
  doi     = {10.1002/cctc.202501054},
  note    = {Early view; full author list and volume to be confirmed at the DOI}
}

@article{Li,
  author  = {Z Li and C Soutis and A Gibson},
  title   = { Overview of Microwave {NDT} Techniques for Fibre-Reinforced Polymer Composites},
  journal = {Appl Compos Mater },
  year    = {2024},
  volume={31},
pages = {1907-1932 }
}

\end{document}